\documentclass[draft]{PJ2019}
\usepackage{url} 
\usepackage{lineno}
\usepackage{soul}
\usepackage{listings}
\usepackage{amsfonts}

\usepackage{amsmath}
\usepackage[normalem]{ulem}

\newcommand {\Dp}[2] {\frac{\partial #1}{\partial #2}}
\newcommand {\Ddp}[2] {\frac{{\partial}^2 #1}{\partial {#2}^2}}

\newcommand{\bmn}{{\mathbf{n}}}

\newcommand{\bmu}{{\mathbf{u}}}

\newcommand{\bmx}{{\mathbf{x}}}

\newcommand{\bmF}{{\mathbf{F}}}

\newcommand{\bmI}{{\mathbf{I}}}

\newcommand{\bmL}{{\mathbf{L}}}

\newcommand{\bmpsi}{\boldsymbol{\psi}}

\newcommand{\bmphi}{\boldsymbol{\phi}}

\begin{document}
\title{Information Flow in geophysical systems}
\authors{P.J. van Leeuwen}
\affiliation{}{Department of Atmospheric Science, Colorado State University, Fort Collins, USA}
\correspondingauthor{P.J. van Leeuwen}{peter.vanleeuwen@colostate.edu}


\begin{keypoints}
\item A new framework on information flows for high-dimensional geophysical systems is derived that only needs low-dimensional calculations.
\item Information evolution is distinctly different from the evolution quantities such as energy or entrophy and differs for different variables.
\item Application to the highly nonlinear systems leads to new understanding of the interaction of the nonlinear terms, and of predictability.

\end{keypoints}

\begin{abstract}
We present a new framework for analyzing the evolution of information in geophysical systems. Understanding how information, and its counterpart, uncertainty, propagates is central to predictability studies and has significant implications for applications such as forecast uncertainty quantification and risk management. It also offers valuable insight into the underlying physics of the system. Information propagation is closely linked to causality: how one part of a system influences another, and how some regions remain dynamically isolated. We apply this framework to the one-dimensional, highly nonlinear Kuramoto–Sivashinsky model and to the shallow-water equations, representing a mid-latitude atmospheric strip. Notably, we observe that information can propagate {\em against} the fluid flow, and that different model variables exhibit distinct patterns of information evolution. For example, pressure-related information propagates differently from relative vorticity, reflecting the influence of gravity waves versus balanced flow dynamics. This new framework offers a promising addition to the diagnostic tools available for studying complex dynamical systems.
\end{abstract}

\section*{Plain Language Summary}
We introduce a new diagnostic quantity, information, to deepen our understanding of geophysical systems. Information is defined as the amount of knowledge available about a specific variable at a given location. We develop a general framework to compute and analyze its temporal evolution, with a focus on how information propagates through the system. This perspective provides valuable insights not only into predictability, but also into the underlying physical processes governing system behavior. We illustrate the framework using both a highly nonlinear toy model and a simplified atmospheric model, demonstrating how it connects to established physical understanding while also revealing novel dynamics.

\section{Introduction}

Although the evolution of physical quantities such as energy, momentum, or vorticity is routinely used to understand the dynamics of geophysical systems, the evolution of uncertainty—or, conversely, information—has received considerably less attention. Most prior work has focused on predictability studies \cite{Carnevale1982,Shukla1998,Smith1999} and data assimilation \cite{VanLeeuwen1996,Evensen2022}, where the growth of uncertainty from initial conditions, model parameters, or boundary conditions is examined to determine quantities such as error doubling times and prediction horizons. However, much more can be learned by studying how information propagates through a system. Information flow may reflect known physical processes or even reveal previously unidentified mechanisms, some of which may not be easily captured by traditional prognostic variables.

Information can be defined in various ways. Intuitively, its opposite, uncertainty, is often associated with the "width" of a variable's probability density function (pdf), but even this width can be quantified in different ways. A widely used and general measure is the Shannon differential entropy:

\begin{equation}
H_{\psi} = - \int p(\psi) \log p(\psi); d \psi
\end{equation}

This definition does not assume a particular shape for the pdf and is applicable to smooth distributions. Generalizations include the Rényi entropies \cite{Renyi1961}, which offer tunable sensitivity to tail behavior and other distributional characteristics.

Entropy-based measures were first introduced into predictability theory by \citeA{Kleeman2002}, who used relative entropy (Kullback–Leibler divergence) to quantify forecast skill relative to climatology. In data assimilation, \citeA{VanLeeuwen2003} used entropy to understand the influence of nonlinear data assimilation in highly nonlinear systems. He found that, contrary to linear data assimilation, in nonlinear data assimilation the posterior variance can be {\em larger} than the prior variance. However, the entropy was still decreasing. 

Entropy has also been central to the field of causal discovery. The causal influence of a set of variables $\bmpsi_e$ on a target variable $\psi$ is typically defined as the portion of the variability in $\psi(t)$ that can be attributed to changes in $\bmpsi_e$ at earlier times, a concept dating back to \citeA{Wiener1956} and \citeA{Granger1969}. While initially applied mainly as lagged linear regression, nonlinear extensions based on information measures, such as mutual information and transfer entropy have been developed over the years, see e.g. the reviews \citeA{Runge2015,Runge2019}. The standard way to display the resulting causal webs is via so-called directed acyclic graphs (DAGs), e.g. \citeA{Pearl2009}, in which arrows from one variable point towards other variables that it influences, quantified in various ways, e.g., via mutual information. DAGs are acyclic by their very nature, excluding cyclic processes by design, reducing their usefulness in real applications. As soon as we redefine causality as the flow of information from a driver variable to a target variable, cyclic processes can easily be included, as shown by \cite{Liang2021} for linear systems. Furthermore, DAGs exclude {\em combined} influences of two or more variables on other variables, as is the typical causal interaction in nonlinear systems. A recently developed causal framework by \citeA{VanLeeuwen2021} addresses the limitation of pairwise-only influence by incorporating interaction information to capture joint effects among variables. While interaction information is often viewed as difficult to interpret, this work demonstrates that, in the context of dynamical systems, it provides a rigorous and physically meaningful measure of joint causality.

While causal discovery has focused primarily on observational data, it can also be applied to numerical models. Although the causal structure of the model equations is explicitly known over a single time step, nonlinear dynamics can obscure causal pathways over longer time horizons. One approach is to freeze the driver variable and observe the resulting change in the target variable. However, such interventions are often problematic. For example, attempting to assess the impact of atmospheric pressure on wind speed by artificially fixing pressure disrupts other feedbacks, thereby invalidating the experiment. Thus, traditional intervention-based methods often fail when applied to prognostic variables in complex, coupled systems. 

A major advance came from connecting dynamical system equations to the evolution of the corresponding probability density function (pdf) via the Fokker–Planck (or Kolmogorov) equation \cite{Fokker1914,Planck1917,Kolmogorov1931}. \citeA{Liang2005} derived expressions for information transfer between variables using this formalism, later extended to multivariate systems \cite{Liang2007a,Liang2007b}, and derived rigorously using the underlying dynamics, and adding explicit links to causality in \cite{Liang2016}, and pushed further for linear systems in \cite{Liang2021}. The theory has been formalized further in \citeA{Pires2024}. These works provided important insights into entropy flows in both continuous and discrete systems. 

Although these papers provided large steps forward, several issues with the formalism remain. Shannon entropy is not invariant under nonlinear
transformations, making the information flows dependent on the variables used. For instance, different results will be obtained when using cloud liquid water content, or its logarithm. Furthermore, the formulation for continuous variables, which is of interest for the present paper, has to deal with the issue that the entropy can be negative, depending on the width of the pdf. As an example, the Shannon entropy of a Gaussian pdf with standard deviation $\sigma$ is equal to $1/2\;\log (2 \pi e \sigma^2)$, which is negative for $\sigma^2 < 1/{2 \pi e}$. This will lead to a sudden change in the direction of the flow of entropy when $\sigma$ crosses this threshold. These issues have been identified by \citeA{Liang2018}, who reformulated the \cite{Liang2016} methodology in terms of relative entropy.

Another limitation of the Liang–Kleeman and \cite{Liang2016,Liang2018} formalisms is computational: it requires integrals over the full state space, making it infeasible for high-dimensional models with millions of variables. \citeA{Pires2024} proposed reducing the dimensionality of the system and performing information-transfer calculations on this reduced system, but this necessarily restricts the range of accessible phenomena that can be studied, and it is not apriori clear how information flow is affected by the dimension reduction.

Finally, the \cite{Liang2016} and related formalisms are all based on keeping potential driver variables frozen to infer causal relations. As discussed above, this can lead to practical computational issues, as fixing a variable can lead to unphysical and even unstable simulations. This problem does not occur in linear causality, because in that case explicit solutions are available, as shown in linear examples in \citeA{Liang2016}.

\sout{Finally, as will be shown in Section 2, the division between self-information and information transferred from other variables is based on a heuristic argument from 
\citeA{Liang2005}, lacking a rigorous foundation. This introduces an arbitrariness that can affect interpretation. }

This paper proposed a new formalism of information and information flow that provides solutions to all these issues. It is based on relative entropy, following \citeA{Kleeman2002} and \citeA{Liang2018}, instead of differential entropy, and all integrals remain low dimensional, even in very high-dimensional systems, exploring the local nature of physical connections. Furthermore, the framework shows that the notion of frozen variables is not needed to determine information flow, and hence causality. The derivation builds on previous entropy-based methods but incorporates essential modifications that make the framework more robust, general, and interpretable. It is applicable to realistic high-dimensional systems, without requiring assumptions about variable transformation or pdf shape.

The framework’s utility is demonstrated using two case studies: the Kuramoto–Sivashinsky equation, a prototypical nonlinear system, and a shallow-water model representing mid-latitude atmospheric dynamics. In both cases, the framework reveals new insights into how information propagates and interacts across variables and scales. We also describe how the necessary integrals can be reliably estimated using relatively small ensembles ($\sim 100$ members), ensuring feasibility for operational and research applications.

The paper is organized as follows. In section 2 the framework will be developed and a simple example shows how the information evolution can be analyzed based on the underlying physics of the system. Then the application of the new framework to the Kuramoto-Sivashinsky model \cite{Kuramoto1978,Sivashinsky1977,Sivashinsky1980} is described in section 3, followed by an application to a shallow-water model of the atmosphere in section 4. Section 5 contains conclusions and a discussion of the results. Three appendices detail some proofs, the extension of the formalism to infinite-dimensional spaces such as described by partial differential equations, and practical ways to calculate the necessary integrals.

\section{Methodology}

We start from the evolution equation of a system with state vector $\bmphi\in \Re^n$, as:
\begin{equation}
\Dp{\bmphi}{t} = \bmF(\bmphi,t)
\end{equation}
This equation is simulated by a numerical model, so even for an infinite-dimensional state this discretization leads to a finite dimensional state vector. The corresponding Fokker-Planck (or Kolmogorov) equation is given by:
\begin{equation}
\Dp{p(\bmphi,t)}{t} =- \Dp{}{\bmphi}\Bigl(\bmF(\bmphi,t)p(\bmphi,t) \Bigr)
\end{equation}
and describes the evolution of the probability density function (pdf) of the vector $\bmphi$ over time and space. 


\subsection{Information}

We are interested in the evolution of information of a specific variable at a specific gridpoint, denoted $\psi(t)$. There is no general definition of information, but there is a clear connection with Shannon entropy. Shannon entropy is defined as:
\begin{equation}
H=-\sum_{i=1}^N p_i \log p_i ,
\end{equation}
for discrete variables, in which $p_i$ is the probability of event $i$ occurring. For continuous variables, as encountered in the geosciences, one can define the differential entropy
\begin{equation}
H_{\psi}=-\int p(\psi) \log p(\psi)\; d\psi ,
\end{equation}
in which $p(\psi)$ is the probability density function (pdf) of continuous variable $\psi$. However, as mentioned in the introduction, a problem with this quantity is that it can be negative for narrow pdfs. This means that the usual interpretation of an entropy as a measure of uncertainty becomes troublesome. As a simple example of a pdf with a negative differential entropy, consider a Gaussian pdf, for which the entropy is equal to $1/2 \;\log (2 \pi e \sigma^2)$, where $\sigma$ is the standard deviation of the pdf. For $\sigma^2 < 1/(2 \pi e)$ the argument of the $\log$ is smaller than one, making the entropy negative.

A way to avoid this issue is to use the relative entropy
\begin{equation}
I_{\psi}=\int p(\psi) \log \frac{p(\psi)} {q(\psi}\; d\psi ,
\end{equation}
in which $q(\psi)$ is a reference pdf. This quantity is also known as the Kullback–Leibler divergence, which serves as a measure of the difference between two probability density functions. Strictly speaking, it is not a true distance metric, as it is not symmetric in its two arguments. However, it is always non-negative and equals zero if and only if the two pdfs are identical, since the logarithmic term becomes zero when the ratio between the pdfs is one.

To apply relative entropy meaningfully, a reference density must be specified. Because we are interested in the evolution of information associated with changes in the pdf 
$p(\psi)$, the reference density should remain constant in time. Furthermore, information should increase as our certainty increases, that is, when 
$p(\psi)$ becomes more concentrated (narrower). This implies that the relative entropy should grow as the divergence from the reference density increases.

These considerations motivate a natural and widely accepted choice for the reference density, the climatological pdf, obtained from a sufficiently long time series such that the distribution becomes statistically stationary. In this study, we adopt this climatological pdf as a reference density to compute relative entropy.

\subsection{Information Flow}
We will now derive an evolution equation for information. To this end, we derive an evolution equation for the information at every variable $\psi$ in the system. The first step is to integrate the Fokker-Plank equation over all variables in $\bmphi$ that do not contain $\psi$. That leads to
\begin{equation}
\Dp{p(\psi,t)}{t} =- \Dp{}{\psi}\int p(\bmphi,t) F_{\psi}(\bmphi,t)  \;d\bmphi\backslash\psi - \int_{\partial \bmphi\backslash{\partial \psi}}p(\bmphi,t)\;\bmF(\bmphi,t)\cdot \bmn \; dS.
\end{equation}
in which $\bmF_{\psi}(\bmphi,t) $ denotes the forcing for variable $\psi$. The second term is a boundary term.
Although the boundary term will be important in many applications, we will assume that it vanishes in the following and leave its detailed treatment to a follow-on paper. 

We now use that, physically, over one time step $\psi$ depends only on its local neighbors $\bmpsi_e$. This means that $F_{\psi}(\bmphi,t)=0$ for those variables in $\bmphi$ that are far from $\psi$, so those variables that do not belong to $\bmpsi_e$. If we denote these variables by $\bmphi_{ne}$ we find:
\begin{eqnarray}
\Dp{p(\psi,t)}{t} & = &- \Dp{}{\psi}\int p(\bmphi,t) F_{\psi}(\bmphi,t)  \;d\bmphi\backslash\psi \nonumber \\
& = &- \Dp{}{\psi}\int p(\bmphi_{ne},t|\psi,\bmpsi_e) p(\psi,\bmpsi_e,t) F(\psi,\bmpsi_e,t)  \;d\bmphi_{ne} d\bmpsi_e,\nonumber \\
& = &- \Dp{}{\psi}\int p(\psi,\bmpsi_e,t) F(\psi,\bmpsi_e,t)  \;d\bmpsi_e
\end{eqnarray}
where we omitted the subscript $\psi$ on $F$ because the meaning of $ F(\psi,\bmpsi_e,t)$ is clear from its first argument.

We now multiply by $[1+\log \bigl(p(\psi,t)/q(\psi)\bigr)]$ and integrate over $\psi$ to find, for the left-hand side:
\begin{eqnarray}
& &\int \Biggl[1+\log \frac{p(\psi,t)}{q(\psi)}\Biggr]\Dp{p(\psi,t)}{t} \; d \psi = \nonumber \\
 & & \qquad \qquad= \Dp{}{t} \int \Biggl[1+\log \frac{p(\psi,t)}{q(\psi)}\Biggr]p(\psi,t) \; d \psi - \int p(\psi,t) \Dp{\log p(\psi,t)}{t} \; d \psi = \nonumber \\
&  & \qquad \qquad=
 \Dp{}{t} \int p(\psi,t) \log\frac{p(\psi,t)}{q(\psi)} \; d \psi = \Dp{I_{\psi}}{t} ,
\end{eqnarray}
where we recognize the last term as the time derivative of the relative entropy of variable $\psi$, and we used 
\begin{equation}
\int p(\psi,t) \Dp{\log p(\psi,t)}{t} \; d \psi = \int \Dp{p(\psi,t)}{t} \; d \psi = \Dp{}{t}\int p(\psi,t) \; d \psi =0.
\end{equation}
because the pdf of a variable integrates to the constant one.

For the first term on the right-hand side we find:
\begin{eqnarray}
& & -\int \Biggl[1+\log \frac{p(\psi,t)}{q(\psi)}\Biggr]
\Dp{}{\psi} \Bigl[F(\psi,\bmpsi_e,t)p(\psi,\bmpsi_e,t) \Bigr] \;d\bmpsi_e d\psi  =  \nonumber \\
& &\qquad \qquad = \int F(\psi,\bmpsi_e,t)p(\psi,\bmpsi_e,t)\Dp{}{\psi}\Biggl( \log \frac{p(\psi,t)}{q(\psi)} \Biggr) \;d\bmpsi_e d\psi ,
\end{eqnarray}
where we again assume that the boundary terms are vanishing.

Putting everything together we obtain:
\begin{equation}
\frac{\partial I_{\psi}}{\partial t}  = 
\int F(\psi,\bmpsi_e,t)p(\psi,\bmpsi_e,t)\Dp{}{\psi} \Biggl(\log \frac{p(\psi,t)}{q(\psi)} \Biggr) \;d\bmpsi_e d\psi,
\label{eq:fullequation}
\end{equation} 
in which $\bmpsi_e$ contains all other variables in the discretized system that are needed to calculate $F(\psi,\bmpsi_e,t)$ over one time step. 

A key feature of the proposed derivation is that all integrals involved are 
low-dimensional. In particular, the highest-dimensional integral arises from the local stencil used to compute the right-hand side of the evolution equation for the variable $\psi$. For complex numerical advection schemes, especially those on unstructured three-dimensional grids, this stencil may still involve a relatively large number of variables. However, a crucial advantage of our approach is that it does not require temporal iteration of the integrals. Instead, we can rely on the underlying numerical model to provide the necessary state variables at each time step. This allows us to use simplified derivative approximations without compromising accuracy. The following section presents a practical example, illustrating how the individual terms in the framework can be computed efficiently.

\subsection{Advection-diffusion equation Example}
As mentioned above, $F(\psi,\bmpsi_e,t)$ denotes the advection velocity of the pdf in $\psi$ space. It typically contains physical advection and diffusion terms, and local terms related to sources and sinks and processes such as phase changes, which only depend on variables in the same model grid point. 

To ease notation we will omit the explicit time dependence in this section. For a continuous system the advection diffusion equation for some quantity $\rho$ reads:
\begin{equation}
\Dp{\rho}{t} = -u\Dp{\rho}{x} - v\Dp{\rho}{y} + D \left(\Ddp{}{x} +\Ddp{}{y} \right)  \rho + \lambda f(\rho,\bmpsi_e) ,
\end{equation}
Our quantity $\psi$ is discretized in space, and hence its spatial derivatives do not exist.Two possible ways forward exist. Either we discretize this equation first, and then employ the framework from section 2.2, or we marginalize the corresponding Liouville equation for the evolution of the measure $\mu_t(\rho)$ in infinite-dimensional space, as detailed in Appendix B. As shown in that appendix, this marginalization still leads to a local description of the pdf of variable $\psi$ at each position in space. If we then discretize the resulting equations, needed for practical evaluation, we find that the two ways forward lead to similar results. Here, we discretize the equation for $\rho$ first, and directly apply the framework from section 2.2, as this will be more familiar for JAMES readers.

We denote the spatial difference operator in the $x$-direction as $L_x(\psi)$, and similar for second-order difference operators. We then find for $F(\psi,\bmpsi_e)$:
\begin{equation}
F(\psi,\bmpsi_e) = -uL_x(\psi) - vL_y\psi) + D \bigl(L_{xx}(\psi) + L_{yy}(\psi) \bigr) + \lambda f(\psi,\bmpsi_e) ,
\end{equation}
in which $u=u(\psi,\bmpsi_e)$, $v=v(\psi,\bmpsi_e)$, and $f(\psi,\bmpsi_e)$ are functions that only depend on the local variables $\psi,\bmpsi_e$. We will evaluate each term below. 

For the $u$ advection we find from Eq. (\ref{eq:fullequation}), suppressing the explicit $t$ dependence:
\begin{equation}
-\int p(\psi,\bmpsi_e) uL_x(\psi)  \Dp{}{\psi}\Biggl( \log \frac{p(\psi)}{q(\psi)}\Biggr) \;d\psi d\bmpsi_e.
\end{equation}
To evaluate this term, we use that the pdfs only depend on $x$ via $\psi$, leading to:
\begin{equation}
L_x\Biggl(\log \frac{p(\psi)}{q(\psi)}\Biggr) \approx 
L_x(\psi) \Dp{}{\psi}\Biggl( \log \frac{p(\psi)}{q(\psi)}\Biggr),
\end{equation}
with an error of order $\Delta x$ or smaller, dependent on the discretization scheme used, as explained in Appendix A.

We then find:
\begin{eqnarray}
& & -\int p(\psi,\bmpsi_e) \;u\;L_x(\psi)  \Dp{}{\psi}\Biggl( \log \frac{p(\psi)}{q(\psi)}\Biggr) \;d\psi d\bmpsi_e = \nonumber \\
& & \qquad \approx   -\int p(\psi,\bmpsi_e) \;u\; L_x\Biggl(\log \frac{p(\psi)}{q(\psi)}\Biggr) \;d\psi d\bmpsi_e = \nonumber \\
& & \qquad \approx -L_x \Biggl(\int p(\psi,\bmpsi_e) u \log \frac{p(\psi)}{q(\psi)} \;d\psi d\bmpsi_e \Biggr)+
\int p(\psi,\bmpsi_e)  L_x(u)\log \frac{p(\psi)}{q(\psi)} \;d\psi d\bmpsi_e .
\end{eqnarray}
The term related to the spatial derivative of $p(\psi,\bmpsi_e)$ does not appear and the error depends on the discretization scheme, as explained in Appendix B.

Similarly, we find for the meridional advection:
\begin{eqnarray}
& & -\int p(\psi,\bmpsi_e) \;v\; L_y(\psi)   \Dp{}{\psi} \Biggl(\log \frac{p(\psi)}{q(\psi)}\Biggr)\;d\psi d\bmpsi_e = \nonumber \\
& & \qquad = -L_y\Biggl( \int p(\psi,\bmpsi_e) v \log \frac{p(\psi)}{q(\psi)} \;d\psi d\bmpsi_e \Biggr) +
\int p(\psi,\bmpsi_e)  L_y(v)\log \frac{p(\psi)}{q(\psi)} \;d\psi d \bmpsi_e .
\end{eqnarray}
We now introduce the {\em advective information flow} as
\begin{equation}
\bmI_{adv} = 
\begin{pmatrix}
\int p(\psi,\bmpsi_e) u \log \frac{p(\psi)}{q(\psi)} \;d\psi d\bmpsi_e\\
\\
\int p(\psi,\bmpsi_e) v \log \frac{p(\psi)}{q(\psi)} \;d\psi d\bmpsi_e
\end{pmatrix} ,
\label{eq:informationflow}
\end{equation}
and an {\em information velocity} via
\begin{equation}
\bmu_I = \bmI_{adv}/I_{\psi} \qquad {\rm such \;that}\qquad \bmI_{adv} = \bmu_{I}I_{\psi}
\end{equation}
The information flow allows us to write for the advection term:
\begin{eqnarray}\label{eq:dH/dtadv}
\left(\frac{\partial I_{\psi}}{\partial t}\right)_{adv} & = &
 \int p(\psi,\bmpsi_e) F_{adv}(\psi,\psi_e)  \Dp{}{\psi} \Biggl(\log \frac{p(\psi)}{q(\psi)} \Biggr)d\psi\;d\bmpsi_e = \nonumber \\
& = & - \bmL_{\bmx} \cdot \bmI_{adv} + 
\int p(\psi,\bmpsi_e) \bigl( L_x(u) + L_y(v) \bigr)\log \frac{p(\psi)}{q(\psi)} \;d\psi d\bmpsi_e ,
\end{eqnarray}
where we use $\bmL_{\bmx}=(L_x,L_y)^T$, in which both components are both spatial difference operators.

For the diffusion term we find the following. First, we write $L_{xx}(\psi) = L_x\bigl( L_x(\psi)\bigr)$. Then, starting with the zonal diffusion term we have:
\begin{eqnarray}
& & \int p(\psi,\bmpsi_e)\; D \; L_{xx}(\psi)  \Dp{}{\psi} \Biggl(\log \frac{p(\psi)}{q(\psi)}\Biggr) \;d\psi d\bmpsi_e = \nonumber \\
& & \;\; = DL_x \Biggl( \int p(\psi,\bmpsi_e)  L_x\Biggl(\log\frac{p(\psi)}{q(\psi)} \Biggr)\;d\psi d\bmpsi_e \Biggr) \nonumber \\
& & \qquad \qquad- D\int p(\psi,\bmpsi_e)  L_x(\psi)L_x\Biggl( \Dp{}{\psi} \Biggl(\log \frac{p(\psi)}{q(\psi)} \Biggr)\Biggr)\;d\psi d\bmpsi_e = \nonumber \\
& & \;\; = D L_{xx}\Biggl( \int p(\psi) \log\frac{p(\psi)}{q(\psi)} \;d\psi \Biggr)  - D\int p(\psi,\bmpsi_e) \bigl(L_x(\psi)\bigr)^2  \Ddp{}{\psi}\Biggl( \log \frac{p(\psi)}{q(\psi)}\Biggr)\;d\psi d\bmpsi_e .
\end{eqnarray}
The meridional diffusion term transforms similarly, giving for the total diffusion term:
\begin{eqnarray}
& & \left(\frac{\partial I_{\psi}}{\partial t}\right)_{diff}=  \int p(\psi,\bmpsi_e) D\left(L_{xx}(\psi) + L_{yy}(\psi)\right) \Dp{}{\psi} \Biggl(\log \frac{p(\psi)}{q(\psi)} \Biggr)\;d\psi d\bmpsi_e = \nonumber \\
& & \;\;  = D\left(L_{xx}+L_{yy}\right) \Biggl(\int p(\psi) \log \frac{p(\psi)}{q(\psi)} \;d\psi \Biggr) \\ \nonumber
& & \qquad - D\int p(\psi,\bmpsi_e)  \left[\left(L_x(\psi)\right)^2 + \left(L_y(\psi)\right)^2 \right]\Ddp{}{\psi}\Biggl( \log \frac{p(\psi)}{q(\psi)}\Biggr) \;d\psi d\bmpsi_e = \nonumber \\
& & \;\;= D\;L_{\bmx\bmx}(I_{\psi}) 
-D \int p(\psi,\bmpsi_e)  \left[\left(L_x(\psi)\right)^2 + \left(L_y(\psi)\right)^2 \right]\Ddp{}{\psi} \Biggl(\log \frac{p(\psi)}{q(\psi)}\Biggr) \;d\psi d\bmpsi_e .
\label{eq:diffusion}
\end{eqnarray}
Finally, for the term without derivatives we can evaluate the integral directly. If $f$ is only a function of $\psi$ we can evaluate further as:
\begin{eqnarray}
& &  \int p(\psi,\bmpsi_e) \lambda f(\psi)  \Dp{}{\psi}\Biggl( \log \frac{p(\psi)}{q(\psi)} \Biggr)\;d\psi\;d\bmpsi_e =  \int p(\psi) \lambda f(\psi)  \Dp{}{\psi} \Biggl(\log \frac{p(\psi)}{q(\psi)}\Biggr)d\psi = \nonumber \\
& &\qquad \qquad = -\lambda \int  p(\psi)\log \frac{p(\psi)}{q(\psi)}  \frac{d f(\psi)}{d\psi} d\psi .
\end{eqnarray}

Combining all the terms we find:
\begin{equation}
\Dp{I_{\psi}}{t} = -\bmL_{\bmx} \cdot \bmI_{adv} + D\;L_{\bmx\bmx}(I_{\psi})  + \left( \Dp{I_{\psi}}{t}\right)_{local} ,
\end{equation}
in which 
\begin{eqnarray}
\left( \Dp{I_{\psi}}{t}\right)_{local} & = & \int p(\psi,\bmpsi_e) \left( L_x(u)+L_y(v)\right)\log \frac{p(\psi)}{q(\psi)} \;d\psi d\bmpsi_e  \nonumber \\ 
& &  - D\int p(\psi,\bmpsi_e)  \left[\left(L_x(\psi)\right)^2 + \left(L_y(\psi)\right)^2 \right]\Ddp{}{\psi} \Biggl(\log \frac{p(\psi)}{q(\psi)} \Biggr)\;d\psi d\bmpsi_e  \nonumber \\ 
& &  + \int p(\psi,\bmpsi_e) \lambda f(\psi,\bmpsi_e)  \Dp{}{\psi} \Biggl(\log \frac{p(\psi)}{q(\psi)}\Biggr)d\psi\;d\bmpsi_e ,
\end{eqnarray}
where, as mentioned, the last term can be simplified if $f$ does not depend on $\bmpsi_e$. The subscript {\em local} denotes that the integrals contain $\log p(\psi)/q(\psi)$ but cannot be evaluated further to an operator acting on relative entropy. This local contribution can be interpreted as information changes due to (1) flow convergence, this term can be positive or negative (2) a modified dissipation term, and (3) the local forcing times the divergence of $\log p(\psi)/q(\psi)$ in $\psi$ space.

We can also write this result as
\begin{equation}
\Dp{I_{\psi}}{t} = -\bmL_{\bmx} \cdot \bmI + \left( \Dp{I_{\psi}}{t}\right)_{local} ,
\end{equation}
in which the advective plus diffusive flux is written as
\begin{equation}
\bmI = \bmI_{adv} + D\;\bmL_{\bmx} (I_{\psi}).
\end{equation}

The local contribution to the change in information, as derived in this work, is not equivalent to the so-called self-information term used in \citeA{Liang2005} and \citeA{Pires2024}. These studies define self-information as:
\begin{equation}
\int p(\psi,\bmpsi_e)\Dp{}{\psi} F(\psi,\bmpsi_e)\;d\psi d\bmpsi_e
\label{eq:self}
\end{equation}
This expression emerges when information is defined via differential entropy rather than relative entropy, and is heuristically interpreted as the change in information of $\psi$ when all other variables, $\bmpsi_e$, are held fixed. That is, when $\bmpsi_e = \bmpsi_{fixed}$. However, while intuitive, this definition proves less useful and even problematic in practice.

To illustrate this, consider the advective-diffusive example discussed above, under the assumption that the velocity components 
$u$ and $v$ are independent of $\psi$ and $\bmpsi_e$. In this case, the self-information term, interpreted as the information evolution of $\psi$ with $\bmpsi_e=\bmpsi_{fixed}$ would yield:
\begin{eqnarray}
\left( \Dp{I_{\psi}}{t}\right)_{local} & = & \int p(\psi,\bmpsi_{fixed}) \left( L_x(u)+L_y(v)\right)\log \frac{p(\psi)}{q(\psi)} \;d\psi  \nonumber \\ 
& &  + \int p(\psi,\bmpsi_{fixed}) \lambda f(\psi,\bmpsi_{fixed})  \Dp{}{\psi} \Biggl(\log \frac{p(\psi)}{q(\psi)}\Biggr)d\psi,
\end{eqnarray}
where the diffusion term is omitted as it cannot be written as a local self information term.
By performing partial integrations and assuming zero boundary terms the last term can be evaluated further, giving:
\begin{eqnarray}
\left( \Dp{I_{\psi}}{t}\right)_{local} & = & \int p(\psi,\bmpsi_{fixed}) \left( L_x(u)+L_y(v)\right)\log \frac{p(\psi)}{q(\psi)} \;d\psi  \nonumber \\ 
&  &   
- \int p(\psi) \lambda\Dp{}{\psi} \Biggl(\frac{p(\psi,\bmpsi_{fixed})}{p(\psi)}f(\psi,\bmpsi_{fixed})\Biggr)\;d\psi 
\nonumber \\
& & 
- \int p(\psi,\bmpsi_e) \lambda f(\psi,\bmpsi_{fixed}) \Dp{}{\psi} \log q(\psi) \;d\psi \nonumber \\
& = & 
 \int p(\psi,\bmpsi_{fixed}) \left( L_x(u)+L_y(v)\right)\log \frac{p(\psi)}{q(\psi)} \;d\psi \nonumber \\ 
&  &   
- \int p(\psi) \lambda f(\psi,\bmpsi_{fixed})\Dp{}{\psi} \Biggl(\frac{p(\psi,\bmpsi_{fixed})}{p(\psi)}\Biggr)\;d\psi  \nonumber \\ 
&  &   
- \int p(\psi,\bmpsi_{fixed}) \lambda\Dp{}{\psi} f(\psi,\bmpsi_{fixed})\Biggr)\;d\psi 
\nonumber \\
& & 
- \int p(\psi,\bmpsi_e) \lambda f(\psi,\bmpsi_{fixed}) \Dp{}{\psi} \log q(\psi) \;d\psi. \nonumber \\
\end{eqnarray}
To compare with the formulation in \citeA{Liang2005} we now adopt differential entropy instead of relative entropy, yielding:
\begin{eqnarray}
\left( \Dp{I_{\psi}}{t}\right)_{local} & = & 
 \int p(\psi,\bmpsi_{fixed}) \left( L_x(u)+L_y(v)\right)\log p(\psi) \;d\psi \nonumber \\ 
&  &   
- \int p(\psi) \lambda f(\psi,\bmpsi_{fixed})\Dp{}{\psi} \Biggl(\frac{p(\psi,\bmpsi_{fixed})}{p(\psi)}\Biggr)\;d\psi  \nonumber \\ 
&  &   
- \int p(\psi,\bmpsi_{fixed}) \lambda\Dp{}{\psi} f(\psi,\bmpsi_{fixed})\Biggr)\;d\psi 
\end{eqnarray}
In contrast, the formulation in \citeA{Liang2005} would yield only:
\begin{equation}
\int p(\psi,\bmpsi_{fixed})\Dp{}{\psi} F(\psi,\bmpsi_{fixed})\;d\psi = 
\int p(\psi,\bmpsi_{fixed})\lambda\Dp{}{\psi} f(\psi,\bmpsi_{fixed})\;d\psi 
\end{equation}
which captures only part of the dynamics relevant to local information flow. 

Moreover, as noted in the introduction, it is often impractical or even impossible to obtain samples from the pdf $p(\psi,\bmpsi_{fixed})$ in complex high-dimensional systems, further limiting the usefulness of this definition in real-world applications.

This example illustrates how the evolution of information can be systematically decomposed into its various contributions. By repeating this procedure for each variable $\psi$ in the system, we can construct an effective information flow field in physical space—or in any relevant variable space. In this introductory study, we refrain from addressing boundary terms associated with the 
$\bmF$ field, which may become important in more realistic, open-domain systems, or systems with bounded variables. These will be the subject of future investigations.

\subsection{Practical ways to calculate the integrals}
Since the shape of the pdfs $p(\psi,\bmpsi_e)$ are not known, a practical way to represent them is via a particle representation. Specifically, we write:
\begin{equation}
p(\psi,\bmpsi_e) \approx \frac{1}{N} \sum_{i=1}^N \delta(\psi-\psi_i)\delta(\bmpsi_e-\bmpsi_{ei}) ,
\end{equation}
and use this expression to change the integrals to summations over particles. This particle-based representation forms the basis for many efficient numerical approximations of information-theoretic quantities.

To approximate the reference density 
$q(\psi)$, several methods can be employed, including kernel density estimation. In the examples presented in this paper, we adopt a Gaussian fit to samples from a long-term model integration as the reference density. This choice is not necessarily an approximation—any arbitrary density can serve as a valid reference—but the Gaussian is shown to be sufficiently accurate for the systems considered here.

Appendix C provides detailed algorithms for accurately evaluating the required summations, including adaptations of existing entropy estimators and newly developed kernel-based estimators for terms such as $d/d\psi_i \log [p(\psi_i)/q(\psi_i)] $, designed to work reliably with relatively small ensemble sizes.

\section{Results on the Kuramoto-Sivashinsky model}
The Kuramoto-Sivashinsky (KS) model is given by:
\begin{equation}
u_t = - uu_x - u_{xx} - u_{xxxx} ,
\end{equation}
in which the subscript denotes differentiation to the subscript variable. The discretized version reads:
\begin{equation}
u_t = - u\;L_x(u) - L_{xx}(u) - L_{xxxx}(u) .
\end{equation}
The model displays nonlinear advection, an anti-diffusion term that tends to enhance gradients, and a stabilizing 4th-order diffusion term to damp small-scale features. The system is chaotic and the route to chaos is via period-doubling bifurcations.

The system is discretized on a regular grid with 1024 gridpoints on a periodic domain, and solved with a pseudo-spectral code. Pdfs are approximated via an ensemble of 200 particles, each starting from a positive cos-squared-shaped bump, as:
\begin{equation}
u_i = A\cos^2\bigl(2\pi(i-525)/100\bigr)
\end{equation}
in position range $[500,550]$ and zero otherwise,
with amplitudes $A \sim N(1,10^{-1})$. This ensemble of initial conditions is propagated forward in time with the KS model equations, and the different terms contributing to the evolution of the information at every grid are calculated. 

\begin{figure}[h!]
\centering
\includegraphics[width=\textwidth]{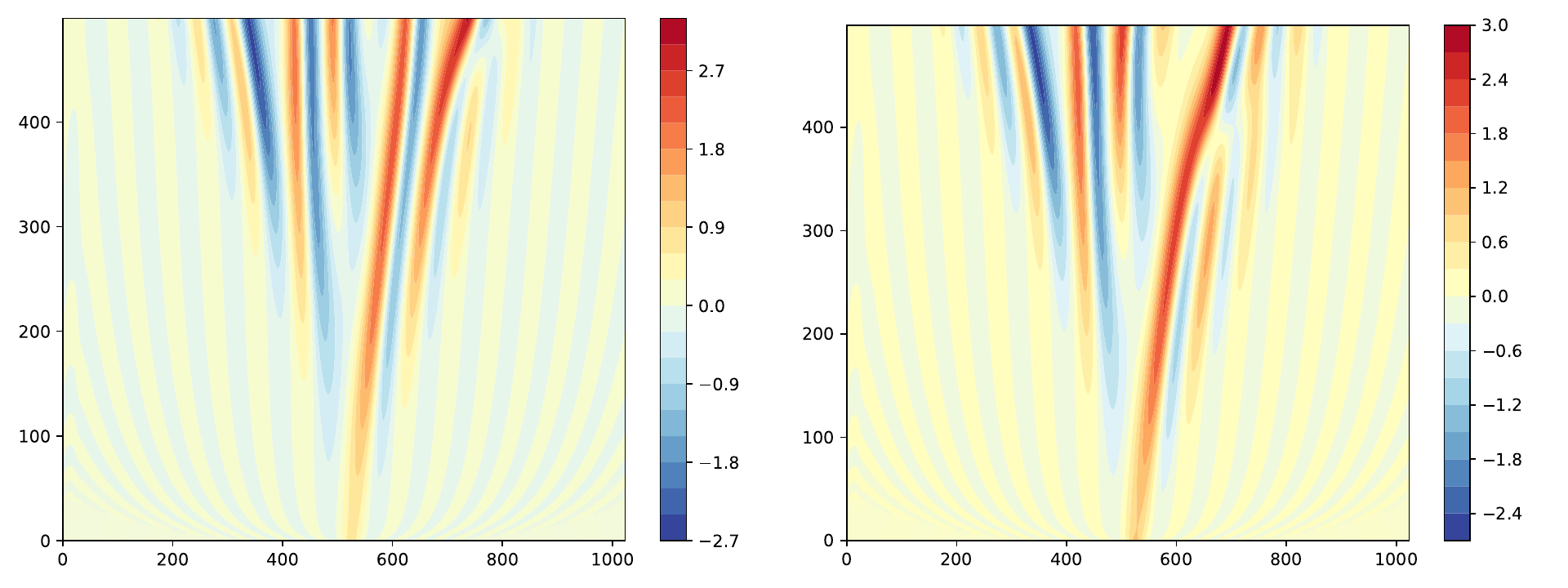}
\caption{Examples of the space-time evolution of two ensemble members. The two members start similar, but deviate substantially at later times.}
\label{fig:KSMembers}
\end{figure}

Fig. \ref{fig:KSMembers} shows the solution for two ensemble members as a space-time plot. Initially, the two members are very similar, and visible differences appear after about 200 time steps, which grow large after 400 time steps. This demonstrates the well-known chaotic nature of the KS equation. The underlying structure is a wave-like feature with wavelength of about 50 grid points, similar to the initial condition for this short simulation.

\begin{figure}[h!]
\centering
\includegraphics[width=\textwidth]{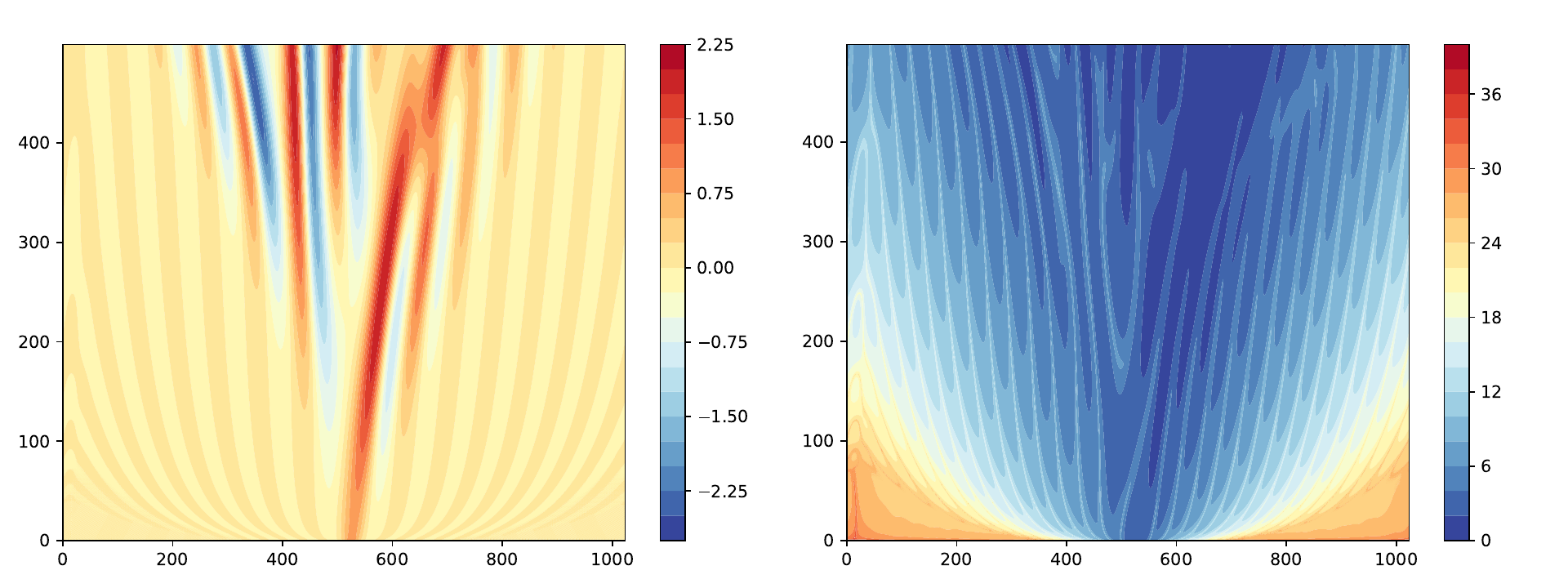}
\caption{Space-time evolution of ensemble mean (left) and Information (right). The horizontal axis is space, and the vertical axis is time. }
\label{fig:MeanAndInformation}
\end{figure}

Figure~\ref{fig:MeanAndInformation} shows the evolution of the ensemble mean 
(left panel) and the information content, defined as the relative entropy or 
Kullback–Leibler divergence, in the right panel. Orange-red colors indicate high information, meaning that the ensemble members are all quite similar, while blue colors show low information. Between time steps 350 and 500, 
the ensemble mean becomes less sharp in the region between grid points 600 and 800. 
This indicates that different ensemble members follow increasingly divergent 
trajectories over time.

The right panel shows that information is initially high, reflecting the tight 
clustering of ensemble members and our high confidence in the state of the system. 
As time progresses, information decreases, indicating reduced confidence and 
increased uncertainty in the state estimate.

An interesting feature is the spatially periodic structure of the information 
field, which reflects the wave-like nature of the KS solution. High information 
values tend to occur near zero crossings of the solution $u$, suggesting that 
ensemble members agree more strongly on the phase (or position) of the waves 
than on their amplitude. This results in the propagation of narrow bands of 
high information along the zero-crossing lines.

Finally, information drops to near zero in the same region where the ensemble 
mean becomes diffuse (around time steps 350-500 and grid points 600–800). 
This indicates that the ensemble pdf begins to resemble the reference 
‘climatological’ pdf $q$, as expected in the later stages of the simulation.

\begin{figure}[h!]
\centering
\includegraphics[width=\textwidth]{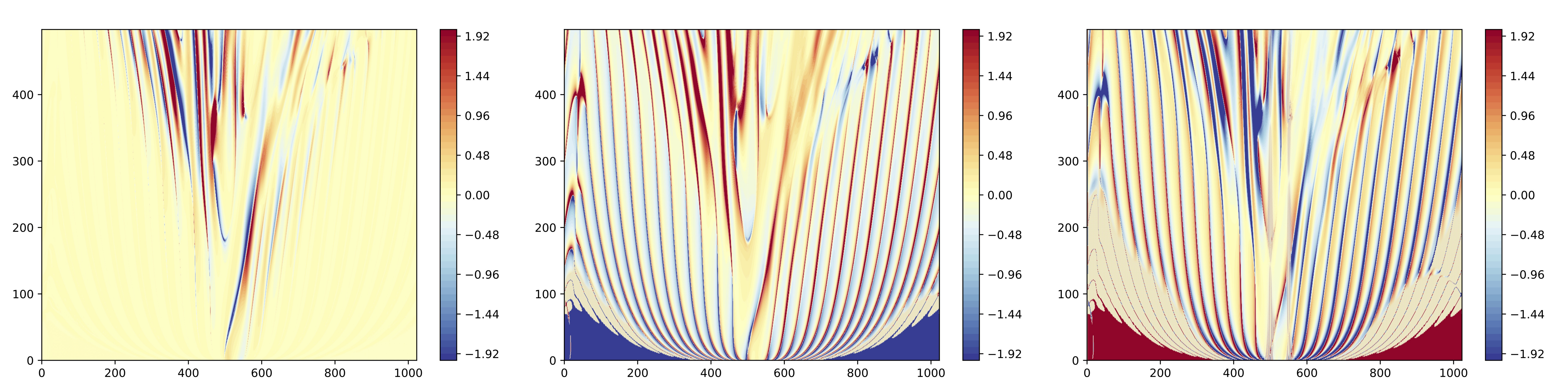}
\caption{Space-time evolution of the forcing terms in the information equation: nonlinear advection (left), anti-diffusion (middle), and dissipation (right).}
\label{fig:KS-InformationSources}
\end{figure}

To better understand the evolution of information, we analyze the forcing terms 
on the right-hand side of the information equation, shown in 
Figure \ref{fig:KS-InformationSources}. These terms correspond to the nonlinear 
advection, the anti-diffusion, and the fourth-order dissipation terms, defined 
as the components in
\begin{equation}
-\int p(u,\bmu_e) \Bigl[ u\;L_x(u) + L_{xx}(u) + L_{xxxx}(u)\Bigr]\Dp{}{u} \Biggl(\log \frac{p(u)}{q(u)}\Biggr)\;du d\bmu_e  ,
\end{equation}
in which $\bmu_e$ contains the $u$ values needed to calculate the 1st, 2nd, and 4th derivatives. Its specific content depends on the numerical scheme used for spatial discretization. Since the numerical model uses a pseudo-spectral method, $\bmu_e$ would contain all grid points. To avoid the resulting high-dimensional integral we used a finite difference representation, using up to 4 grid points $\bmu_e$. While an approximation, this is still quite accurate since we perform this calculation at each time step from the psuedo-spectral solution, such that there is no accumulation of discretization errors over time. 
To better reveal the spatial structure of the terms, the plotted values are clipped to the range $[-2, 2]$. 

One might expect that the nonlinear advection and the anti-diffusion terms would decrease information (hence blue) because they are responsible for instabilities in the flow. Furthermore, the dissipation term is expected to increase information (hence red) because it dissipates small-scale structures. However, the information changes are more subtle. 

Three main regions can be identified: (1) early times and peripheral grid points, 
where broad red and blue areas appear, (2) a widening central region where the 
nonlinear advection term dominates, and (3) the area defined by the smooth curves in between. 

The expected roles of anti-diffusion (information-reducing) and dissipation 
(information-increasing) are observed at early times and away from the central axis, in region (1). 
At later times, especially in the central region (2), this behavior changes. 
The anti-diffusion term becomes positive (increasing information), while the 
dissipation term fluctuates in sign, much like the advection term. In fact, a 
closer inspection reveals that the combined effects of the advection and 
anti-diffusion terms approximately cancel the dissipation term.

Between these two areas in region (3), smooth curves are visible in the anti-diffusion and dissipation terms, similar to the wave-like solution in the solution of the ensemble mean and the information itself in Fig \ref{fig:MeanAndInformation}. Closer inspection shows that in the anti-diffusion plot all curves are red towards the central axis and blue away from this axis, and the dissipation plot shows the colors the other way around. If we assume a sinusoidal solution in space, e.g. $\sin kx$, the anti-diffusion term would be $k^2 \sin kx$ and the dissipation term would be $-k^4 \sin kx$, so the two terms would have opposite sign. However, the solution for the ensemble mean shows a dominant wavelength that is half that of the anti-diffusion and dissipation terms. 

This paradox is solved by realizing that the anti-diffusion and dissipation terms can be written as anti-diffusion and dissipation of the {\em information}, plus local information sources and sinks, as in Eq. (\ref{eq:diffusion}). Information is low at both positive and negative values of the ensemble mean solution of the KS equation, explaining the dominant wavelength shift.

If we follow the evolution of information at a single grid point within this 
region (3), we find that as a high-information wave passes, the anti-diffusion 
term initially decreases information, while dissipation increases it. Later, 
the signs reverse. This results in a net information change resembling the 
dissipation forcing, suggesting that, in this regime, dissipation plays the 
dominant role.

Finally, consider the central region (2), where nonlinear advection is strong and the information field exhibits narrow bands of high values, aligned with zero-crossings of the wave structures. Here too, the forcing terms are most prominent along these curves. The anti-diffusion term tends to increase information (red), while the advection and dissipation terms have both positive and negative contributions that largely balance the former. This more nonlinear regime is harder to understand and the wave-like solution is lost. The nonlinear advection term is much more sensitive to small differences between the ensemble members, and higher values are less confined to the lines, and increases information in some places and decreases it elsewhere. This more random structure in the nonlinear advection term is largely compensated by the dissipation term. We see that the dissipation term acts differently on information than on the state vector $u$, where it will always tend to reduce energy. 

\begin{figure}[h!]
\centering
\includegraphics[width=\textwidth]{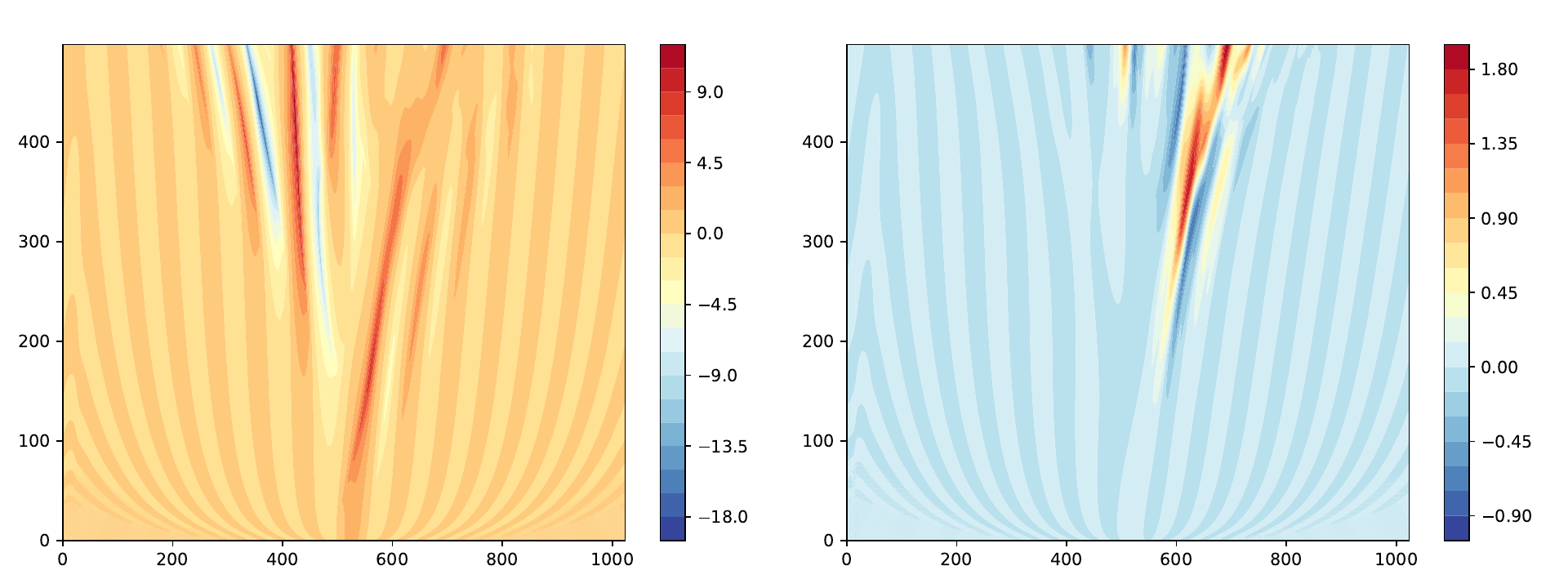}
\caption{Space-time evolution of information flow (left) and difference between information flow and ensemble mean solution times information (right)}
\label{fig:InformationFlow}
\end{figure}

In Fig. \ref{fig:InformationFlow} we display the advective information flow, as defined in Eq. (\ref{eq:informationflow}), on the left panel. Positive values (orange/red) denote flow to the right, and negative values (yellow/blue) flow to the left. Note the alternating flows away from the middle of the plot, related to the wave-like structure of the solution of the KS equation. In the central region we see structures similar to those in the ensemble mean in Fig. \ref{fig:MeanAndInformation}, but with larger amplitudes. This raises the question if the information flow is simply information advected with the mean velocity.

To explore this, we compare the information flow $\bmI_{adv}$ with the advection of the information with the flow, defined as $uI_{u}$, by subtracting the latter from the former. While in most of the domain the information flow is similar to the advection of information, red (positive) and deeper blue (negative) values appear later in the evolution, especially in area where the information decreases rapidly, related to different instabilities in the different ensemble members.

This simple example shows that information is not merely advected with the flow, but can also grow or decay due to local instabilities. It can also be created when the flow stabilizes and ensemble spread decreases. Instead of just following ensemble spread, the framework developed above allows us to understand which terms are driving the grow and decay of information.

\section{Results on the shallow-water model}
This two-dimensional shallow-water model has three prognostic variables: the
two wind components $u$ and $v$, and the depth of the fluid layer $h$. The model simulates a channel around the Earth between 20 and 70$^o$N with a spatial resolution of 100 km (around one degree), and uses a time step of 1 min. The domain is periodic in the $x$ direction and has solid north and south boundaries where $v = 0$, and where $h$ is fixed at its initial values. The shallow water equations in conservative (or flux) form may be written as:
\begin{eqnarray}
\Dp{h}{t} & = & - \Dp{hu}{x} - \Dp{hv}{y} \\ \nonumber
\Dp{hu}{t} & = & - \Dp{}{x}\Bigl(u^2 h + \frac{1}{2} gh^2 \Bigr) -\Dp{}{y} huv + f hv \\ \nonumber
\Dp{hv}{t} & = & -\Dp{}{x} huv- \Dp{}{y}\Bigl(v^2 h + \frac{1}{2} gh^2 \Bigr)  - f hu,
\end{eqnarray}
where $f$ is the Coriolis parameter and $g$ is the acceleration due to gravity. In this model, the Coriolis parameter is modeled as $f= f_0 + \beta(y-y_0)$, a so-called beta-plane approximation, with $y_0$ at the middle of the domain. 

We integrate this model forward-in-time using the Lax-Wendroff scheme, which achieves second-order accuracy and requires only one previous time-step, and does not need any artificial diffusion in time or space to keep it stable. The scheme can be considered a centered-time, centered-space step in two dimensions, exploring half-time steps. The spatially discretized version of the model uses as $L_x$ and $L_y$ operators standard central differencing, and is not displayed here.

The model is initialized with the pressure field from the ECMWF reanalysis from 1 July 2000. This field is actually from the southern hemisphere, but flipped to be run in the northern hemisphere with no orography. The pressure field is converted to the shallow-water variable height via $h = 0.99*pressure/g$. The initial velocity fields are in geostrophic balance with this pressure field, i.e., discretized versions of:
\begin{eqnarray}
fv & = & g\Dp{h}{x}  \\ \nonumber
fu & = & - g\Dp{h}{y} .
\end{eqnarray}

\begin{figure}[h!]
\includegraphics[width=0.95\textwidth]{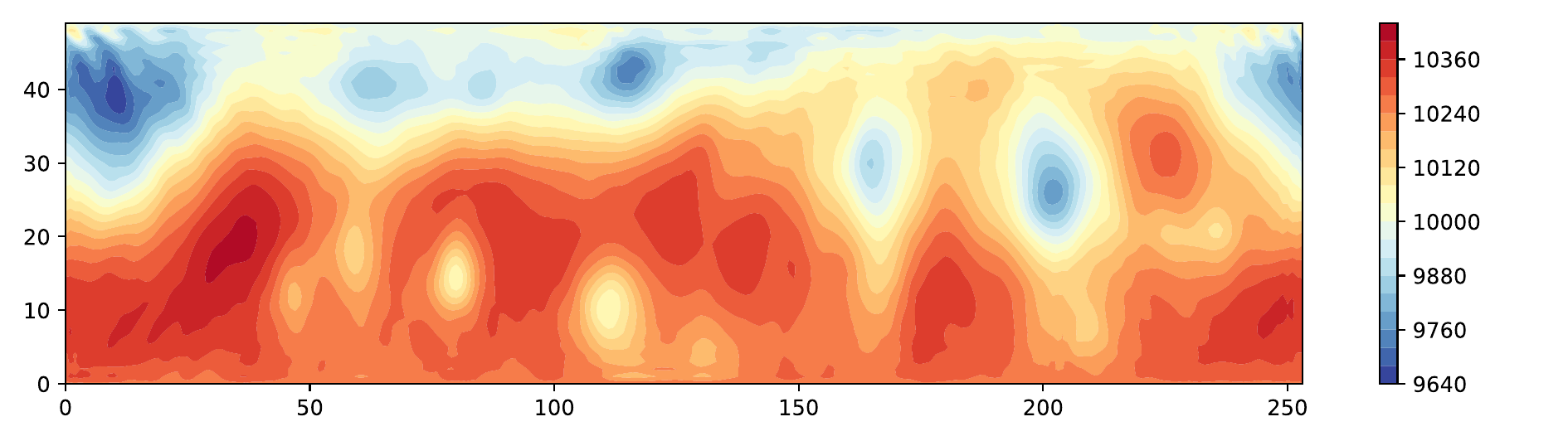}
\caption{Example of the height field during a simulation.}
\label{fig:Pressure}
\end{figure}

The model has 254 gridpoints in the zonal direction and 50 gridpoints in the meridional dimension. Fig. \ref{fig:Pressure} shows an example of the height field of the simulations. The air tends to flow along contours of equal height, and the dominant flow is from west to east, meandering along the middle latitude of the domain. High and low pressure cells are clearly visible too. 

\begin{figure}[h!]
\includegraphics[width=0.9\textwidth]{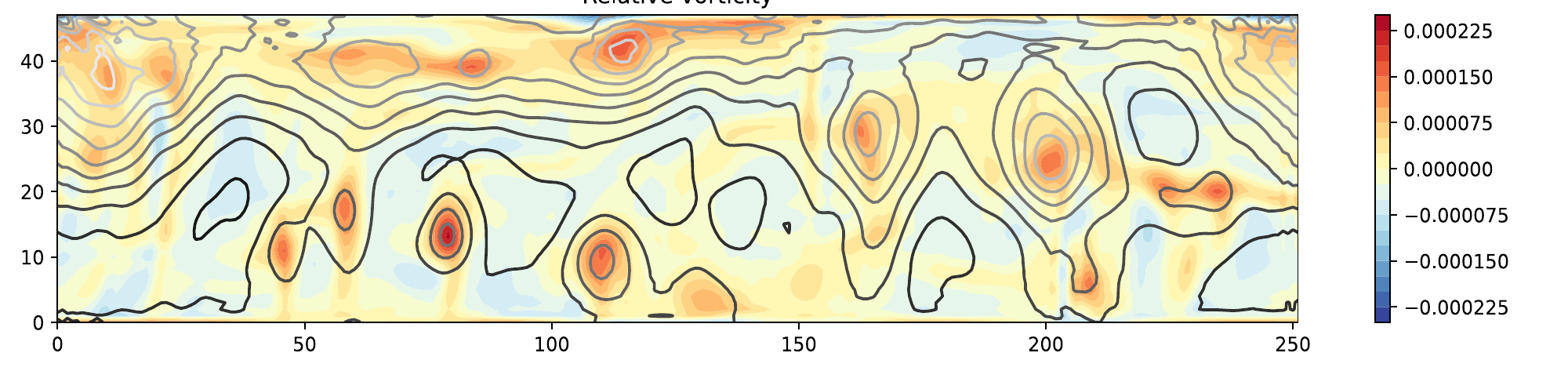}
\caption{Relative vorticity field, with grey contour of equal pressure.}
\label{fig:RelativeVorticity}
\end{figure}

The relative vorticity, defined as $\zeta = v_x-u_y$, is displayed in Fig \ref{fig:RelativeVorticity}. The curves are contours of equal pressure. Comparing to the height field in Fig. \ref{fig:Pressure}, we see that low-pressure areas correspond to high relative vorticity, and vice versa. The Coriolis parameter in the middle of the domain is $10^{-4}\;s^{-1}$, while typical relative vorticity values are twice that magnitude, showing that the dynamics is dominated by nonlinear dynamics, with Rossby numbers of order 2.

\begin{figure}[h!]
\includegraphics[width=\textwidth]{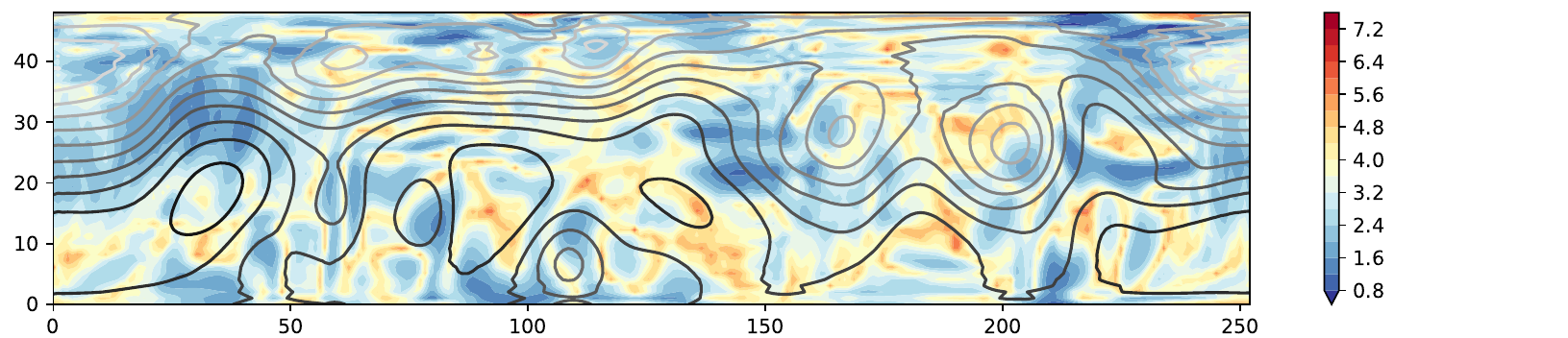}
\caption{Information field of the relative vorticity, overlaid with equal pressure contours.}
\label{fig:SW-InformationZeta}
\end{figure}

An ensemble of 200 model states is generated by running the model and sampling the fields every 10 time steps. Then each sample is run for a day and the information-related quantities are calculated. The reference pdf $q$ is obtained by calculating the mean and variance of the state over time, ensemble and zonal direction. This results in a meridional profile for the mean and the variance, and these are assumed constant in the zonal direction. We choose this procedure because the model is translation invariant in the zonal direction. 

Fig. \ref{fig:SW-InformationZeta} shows the information field of the relative vorticity field, with contours of equal height overlaid. The relative-vorticity equation for this model can obtained from the model equations above as:
\begin{equation}
\Dp{\zeta}{t} = - u\Dp{\zeta}{x} - v\Dp{\zeta}{y} - \beta v - (\zeta+f)\biggl(\Dp{u}{x}+\Dp{v}{y} \biggr) ,
\end{equation}
in which the constant $\beta=\partial f/\partial y$. This equation contains advection of relative vorticity, advection of planetary vorticity (the $\beta v$ term), and a local stretching term. The spatially discretized version is written as 
\begin{equation}
\Dp{\zeta}{t} = - u\;L_x(\zeta) - v\;L_y(\zeta) - \beta v - (\zeta+f)\biggl(L_x(u) + L_y(v) \biggr) ,
\end{equation}
in which $L_x,L_y$ are central differences.

The information field contains small-scale structures which remain when the ensemble size is increased by factors 2 or 4, suggesting that they are real and directly connected to information dynamics. Overall, areas with high information (yellow/red) tend to concentrate at and around centers of high- and low-pressure cells, while low information areas (blue) tend to concentrate in areas of large meridional displacements of the Jet Stream. This is not unexpected, as those areas are indications of flow instability, leading to larger ensemble spread and hence lower information. 

In more detail, the high-information areas tend to be narrow 'streaks' in areas of low air flow (low concentration of pressure contours), but sometimes the air flow is large, as for instance in the region (110-130 E, 30-40 N). This shows that information at a certain time is not only determined by the flow structures at that time, but is rather the result of an evolving flow. Indeed, information cannot be derived from the traditional local flow quantities, but the pdf-information is needed, and that is determined by the evolution of information over time.

\begin{figure}[h!]
\includegraphics[width=0.95\textwidth]{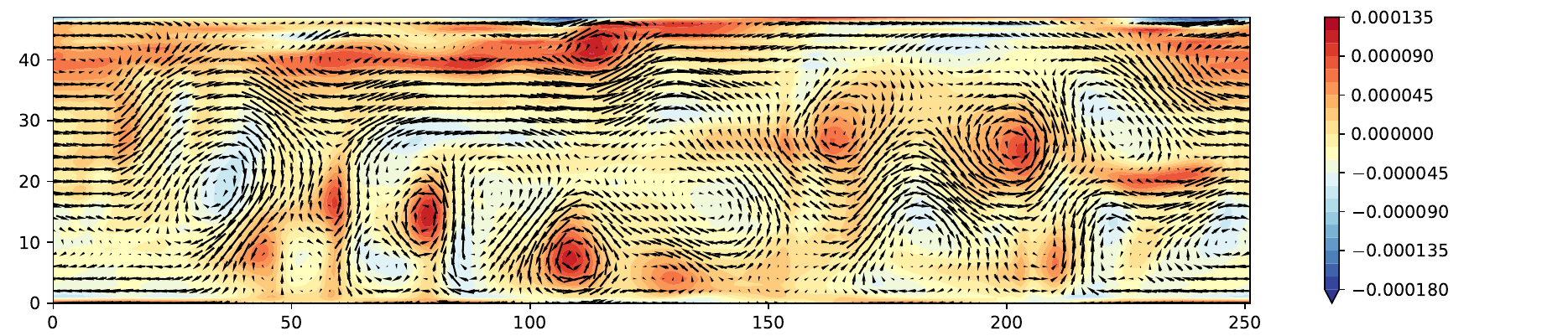}
\caption{Advective Information flow of relative vorticity, overlaid on the ensemble-mean vorticity field.}
\label{fig:SW-InformationFlowZeta}
\end{figure}

\begin{figure}[h!]
\includegraphics[width=0.5\textwidth]{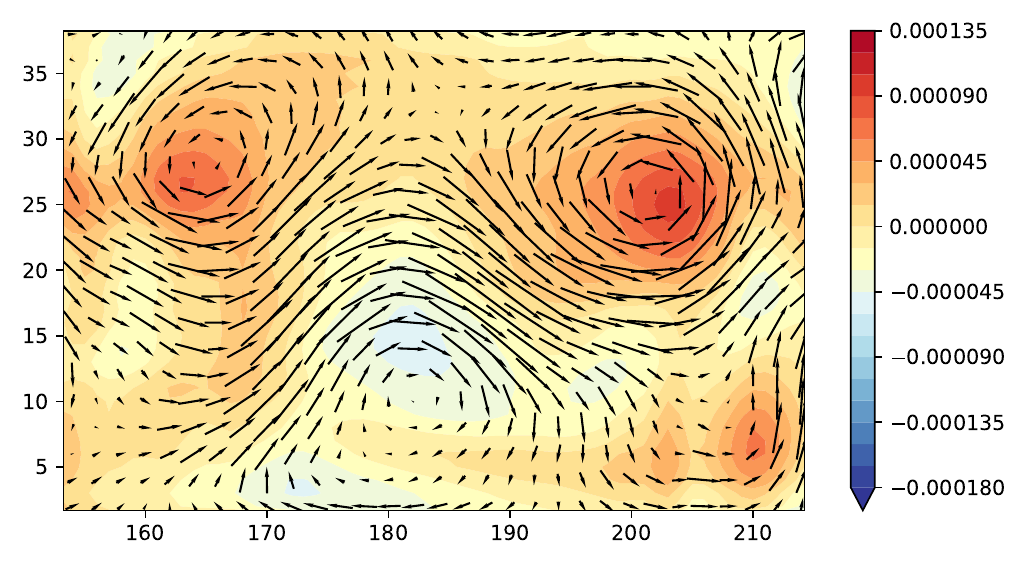}
\caption{Detail of the advective Information flow of relative vorticity, overlaid on the ensemble-mean vorticity field.}
\label{fig:SW-InformationFlowDetail}
\end{figure}

A snap shot of the advective part of the information flow of relative vorticity, as in Eq. (\ref{eq:informationflow}), which, since the model contains no explicit diffusion, can be considered the full information flow, is shown in Fig. \ref{fig:SW-InformationFlowZeta}, overlaid on the relative vorticity field. We see that information flow is largely directed along contours of equal relative vorticity, as expected. However, a close examination, aided by an enlargement in Fig. \ref{fig:SW-InformationFlowDetail}, shows that information-flow direction and magnitude do not align with vorticity. The flow is often displaced, sometimes by more than 5 grid points, or 500 km. This displacement is not systematically in one direction, but seems to be related to the larger-scale flow. 

A comparison of the displacement of the information flow from relative-vorticity contours, or where the information flow crosses the relative-vorticity colors, with the size of the information in Fig. \ref{fig:SW-InformationZeta} might suggest that larger displacements correspond to smaller information values (more blue), but we do not expect such a direct connection.

As a further illustration of the power of the methodology, the information evolution of two variables with different dynamical characteristics is discussed. We initialize the ensemble from the ECMWF pressure field, but add small Gaussian noise $\eta_1$ of amplitude 1 m and spatial correlation lengthscale of 5 gridpoints everywhere, and add a Gaussian shaped random perturbation:
\begin{equation}
dh(i,j) = \eta_1(i,j) + A \eta_2\exp \Bigl[- \frac{(i-ic)^2+(j-jc)^2}{2 L^2}\Bigr] ,
\end{equation}
in which $ic=120$, $jc=30$, $L=4$,and $A=100\;m$ and $\eta_2 \sim N(0,1)$. The initial velocity fields are in geostrophic balance with the ECMWF pressure field, but not with this random perturbation. 

\begin{figure}[h!]
\includegraphics[width=0.95\textwidth]{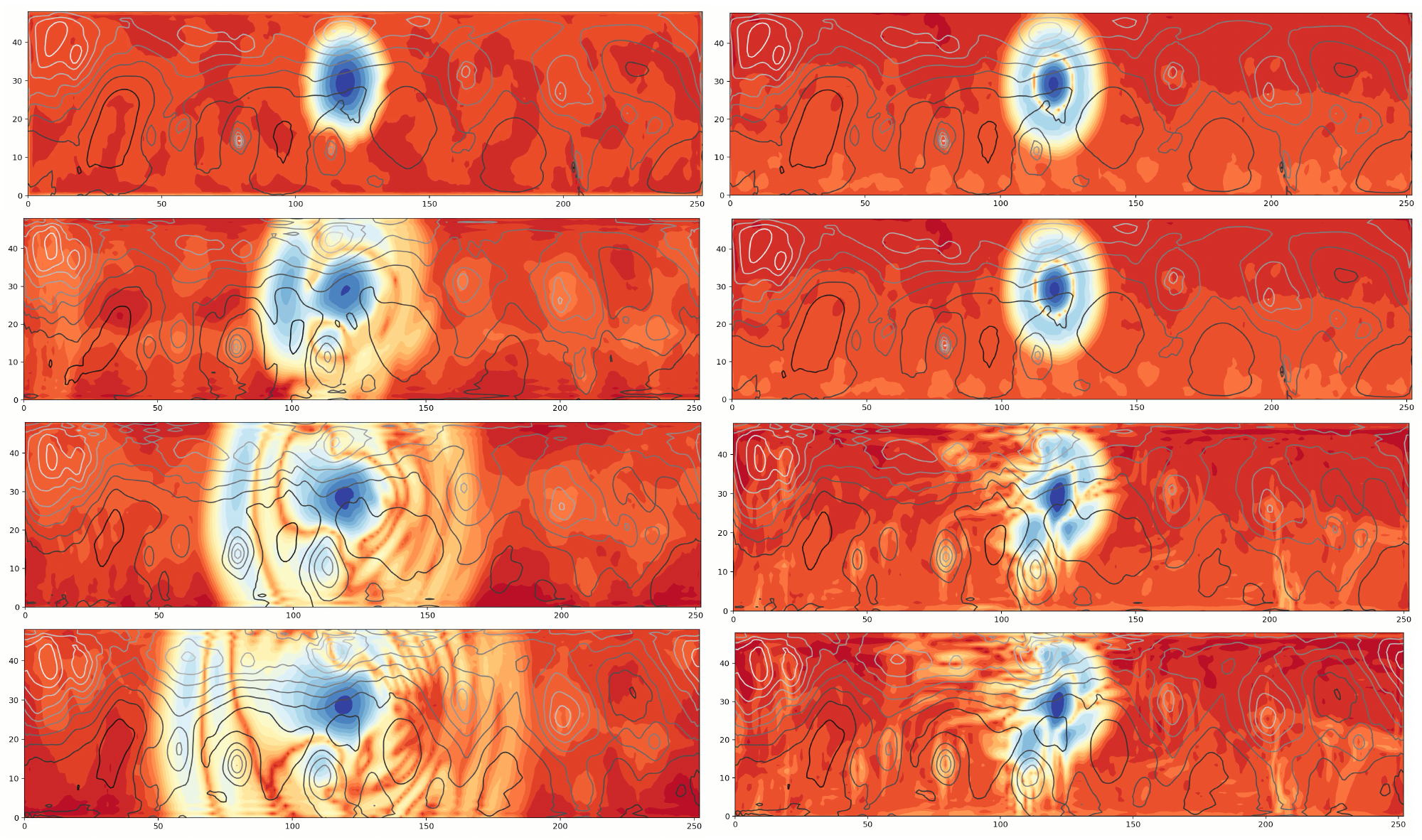}
\caption{Information field evolution of the height (left) and relative vorticity (right), with grey contours of equal background pressure. The time interval is 1 hour and 40 minutes.}
\label{fig:BlobEvolution}
\end{figure}

Fig. \ref{fig:BlobEvolution} shows the evolution of the information of the height field in the left panels, and of the relative vorticity field on the right panels. The time interval between the panels is 1 hour and 40 minutes. 

The striking difference between the left and right panels is the wavelike pattern in the left panels that radiates away from the source region. Its propagation speed is about $390\;m/s$, a clear signal of gravity waves, which have a theoretical propagation speed of $\sqrt{gh}\approx 330\;m/s$. The formation of gravity waves is expected from geostrophic adjustment, in which gravity waves are emitted to facilitate the development of geostrophically balanced flow. Hence, information is propagating with these gravity waves. The wave propagation to the west is larger than the eastward geostrophic flow, resulting in an information propagation against the flow of the air parcels.

Furthermore, while the main flow is toward the East, the information remains concentrated when flowing westward, while it  disperses more flowing eastward. This might be due to a similar effect that concentrates wave energy when waves propagate against a mean flow, but a further investigation is beyond the scope of this paper.

A large information 'blob' remains behind at the source region, showing the geostrophically adjusted information propagation closely aligned with the geostrophic flow, and with the relative vorticity field in the right panel. The reason why the relative vorticity information does not radiate out is that the gravity waves are rotation-free and do not carry relative vorticity. Instead, the relative-vorticity information starts to deform according to the background velocity, i.e. is starts to align with the background pressure contours.

In both the height and relative vorticity information evolution we notice the development of narrow 'streaks', similar to what was found in the original experiment in Fig. \ref{fig:SW-InformationZeta}. The height information evolution suggests that gravity waves might play a larger role in these. In contrast, the relative vorticity information evolution shows the deformation of a high information ring within the low-information 'blob', which might suggests that the structures in Fig. \ref{fig:SW-InformationZeta} are more advective in nature. 

\begin{figure}[h!]
\includegraphics[width=0.9\textwidth]{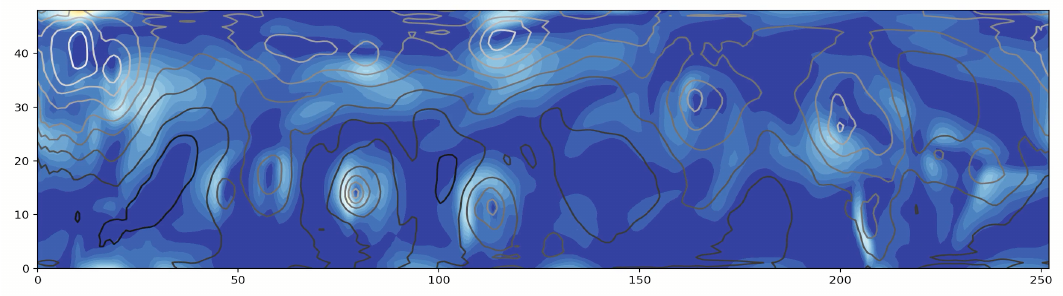}
\caption{Energy field after 1 hour and 40 minutes, with grey contours of equal pressure.}
\label{fig:Energy}
\end{figure}

The above shows that information is not just advected with the air flow, but might suggest that its evolution is more closely related to that of energy, which, in this barotropic model, is mainly kinetic. However, the kinetic energy field, displayed in Fig. \ref{fig:Energy} for 1 hour and 40 minutes into the simulation, does not show this behavior at all. This shows that the information evolution is not directly related to any geophysical quantity. It is the evolution of uncertainty in the system. A figure of the {\em information} of the energy shows a mixture of the behavior of height and relative vorticity, not shown here. Much more can be said about the evolution of information, but given the explorative nature of this paper, we leave that for subsequent work.

\section{Conclusions}
A new framework is presented to calculate and analyze the evolution of information in complex high-dimensional systems. It addresses several limitations of existing approaches, including the reliance on differential entropy, which can be negative, challenges associated with high-dimensional integrals, and ambiguities in the definition of self-information and the use of frozen variables. 

The formalism is formulated in finite-dimensional discrete spaces to allow for the use of probability density functions. However, in Appendix B we also discuss an extension to infinite-dimensional systems, and show that even in those systems the resulting expressions remain local, such that, when the equations are discretized for practical computations, all integrals remain low dimensional.

An initial analytical application to the advection-diffusion equations in two dimensions demonstrates how the information forces can be analyzed and decomposed into other intuitive terms. The {\em information flow} and {\em information velocity} are introduced as central concepts to aid in understanding of information dynamics. 

We apply this formalism to numerical solutions of both the Kuramoto–Sivashinsky and the shallow-water models, and demonstrate how it enables a deeper understanding of system evolution from an information-theoretic perspective. Notably, we show that information flow does not necessarily follow the physical velocity field. Furthermore, information dynamics can be vastly different for different variables in the system, demonstrated in this paper in a geostrophic adjustment experiment.  In general, information does not map directly onto any individual system variable, and its evolution follows dynamics that are distinct from, for instance, energy propagation. 

This framework provides a fundamentally new lens through which to study complex high-dimensional systems. With the growing recognition that the propagation of information, or uncertainty, is as crucial as energy transport in many scientific contexts, information-based diagnostics may become an essential addition to the scientific toolkit. Potential applications include studies of information dynamics in full-Earth system models, high-resolution air–sea interaction studies, and large-eddy simulations of cloud development.

More broadly, the framework offers promising contributions to our understanding of system dynamics, predictability, data assimilation, and risk quantification. This opens the door to a wide range of societal applications, from weather forecasting and environmental monitoring to risk management, insurance, and the development of digital twins. This paper provides only an initial demonstration of its wide application area.

\newpage
\appendix
\section{Spatial derivative approximations}

In the main text several approximations were made to arrive at our final expressions. In this appendix we will discuss the three basic steps we took.

First, we rewrite $L_x(f(\psi)$ in terms of $L_x(\psi)$, in which $f(..)$ is some smooth function of $\psi$. By definition, we can write
\begin{equation}
L_x(f(\psi)) = \sum_i b_i f(\psi_i)
\end{equation}
in which $b_i$ are constants defining the discretization scheme, and $\psi_i$ are values of $\psi$ at neighboring gridoints $x_i$. The $b_i$ are $O(1/\Delta x)$. Since $L_x$ represents a discretized version of a spatial derivative, and the spatial derivative of a constant is zero, we find that if we take $f(\psi)=1$ that $\sum_i b_i = 0$. With this, we can write, using a first-order Taylor expansion:
\begin{eqnarray}
L_x(f(\psi)) & = & \sum_i b_i f(\psi_i)  \nonumber  \\  & = & \sum_i b_i (\psi_i-\psi) \Dp{f(\psi)}{\psi} +  O \Bigl(\sum_i b_i(\psi_i-\psi)^{2}\Bigr) \nonumber  \\  & = & 
\sum_i b_i \psi_i \Dp{f(\psi)}{\psi} +  O\Bigl(L_x(\psi^2)-2\psi L_x(\psi)\Bigr)
\nonumber  \\  & = &
L_x(\psi) \Dp{f(\psi)}{\psi} +  L_{xx}(\psi)O\bigl(\Delta x^n \bigr)
\end{eqnarray}
in which $O(\Delta x^n)$ is the accuracy of the numerical discretization used in the numerical model. In the steps above we used $\sum_i b_i f(\psi) = 0$ and $\sum_i b_i \psi \partial f(\psi)/\partial \psi = 0$, due to $\sum_i b_i = 0$.

Second, we rewrite $f(\psi)\;L_x(g(\psi))$ in terms of $g(\psi)\;L_x(f(\psi))$, where $f(..)$ and $g(..)$ are smooth functions of $\psi$. We have, defining $\Delta f_i = f(\psi_i)-f(\psi)$:
\begin{eqnarray}
L_x\Bigl(g(\psi)f(\psi)\Bigr) & = & \sum_i b_i f(\psi_i) g(\psi_i) = \sum_i b_i (f(\psi) + \Delta f_i)(g(\psi)+\Delta g_i) 
\nonumber  \\  & = & 
\sum_i b_i f(\psi)  (g(\psi)+\Delta g_i) +
\sum_i b_i g(\psi)
(f(\psi) + \Delta f_i) 
+ \sum_i b_i \Delta g_i \Delta f_i
\nonumber  \\  & = & 
f(\psi) \sum_i b_i g(\psi_i) +
g(\psi) \sum_i b_i f(\psi_i) + 
O \Bigl(\sum_i b_i(\psi_i-\psi)^{2}\Bigr)
\nonumber  \\  & = &
f(\psi) L_x(g(\psi)) +
g(\psi) L_x(f(\psi))
+  L_{xx}(\psi)\bigl(\Delta x^n \bigr)
\end{eqnarray}
or
\begin{equation}
f(\psi)\;L_x(g(\psi)) = 
L_x\Bigl(g(\psi)f(\psi)\Bigr) - g(\psi)\;L_x(f(\psi)) +
O(\Delta x^n)
\end{equation}
in which $O(\Delta x^n)$ is the accuracy of the numerical discretization scheme.

Finally, we examine the operation of $L_x$ on probability integrals. We assume that $L_x$ uses $m$ variables $\psi_i$. We can write:
\begin{eqnarray}
L_x \Biggl(\int p(\psi)  g(\psi)\;d\psi \Biggr)& = &
\sum_i b_i  \int p(\psi_i)  g(\psi_i)\;d\psi_i
\nonumber \\
& = &
\sum_i b_i  \int p(\psi_1,...,\psi_m)  g(\psi_i)\;d\psi_1 ...d\psi_m 
\nonumber \\
& = &
\int p(\psi_1,...,\psi_m)  \sum_i b_i g(\psi_i) \;d\psi_1 ...s\psi_m 
\nonumber \\
& = &
\int p(\psi) L_x( g(\psi) )\;d\psi 
.
\end{eqnarray}

\section{The infinite-dimensional case}
Starting from the evolution of the measure describing solutions from a partial differential equation (PDE) in an infinite dimensional space, we will derive an equation for the evolution of the probability density function (pdf) in a finite-dimensional subspace. Details can be found in \citeA{Ambrosio2008}. We then show that the fact that physical systems are causal, i.e., each physical system has a finite fastest speed with which information can propagate through that system, leads to a locally closed-form solution of the evolution of the pdf. With this, we show that the finite-dimensional information evolution discussed in the main text can be extended to the infinite-dimensional domain, i.e., to systems described by PDEs.

\subsection{PDEs and Their Associated Liouville Equation}
Consider a PDE on a Hilbert space $H$:
\begin{equation}
\frac{\partial u}{\partial t} = F(u)
\end{equation}
where $F: H \to H$ is an operator encoding the dynamics. 

Given an initial measure $\mu_0$ on $H$ (representing uncertainty over initial conditions), the measure $\mu_t$ evolves according to the Liouville equation (also called the continuity equation or transport equation):
\begin{equation}
\frac{\partial \mu_t}{\partial t} + \nabla \cdot (\mu_t F) = 0
\end{equation}

The meaning of this equation follows from the weak identity involving test functions $\varphi: H \to \mathbb{R}$ as:
\begin{equation}
\frac{d}{dt} \int_H \varphi(u) \, d\mu_t = \int_H \langle D\varphi(u), F(u) \rangle \, d\mu_t
\end{equation}
in which $D\varphi(u)$ is the Frechet or directional derivative. For any direction $h \in H$, the directional derivative is defined as:
\begin{equation}
\langle D\varphi(u), h \rangle = \frac{d}{ds}\bigg|_{s=0} \varphi(u + sh) 
\end{equation}

\subsection{Projection to Finite Dimensions and Marginalization}
We seek a finite-dimensional description by projecting the infinite-dimensional space $H$ on a finite dimensional space $H_N$, so that we can write:
\begin{equation}
H = H_N \oplus H_\perp
\end{equation}
where $H_N = \text{span}\{e_1, \ldots, e_N\}$ is a finite-dimensional space, and $H_\perp$ is the orthogonal complement. As an example, the finite-dimensional space might be the variable $\psi$ at a certain grid point, as in the main text.

We now define a {\em cylindrical} test function via
\begin{equation}
\varphi(u) = f(\pi(u))
\end{equation}
where $\pi: H \to \mathbb{R}^N$ is the projection onto $H_N$, so the vector of coordinates of its argument with respect to the basis $\{e_1, \ldots, e_N\}$, and $f: \mathbb{R}^N \to \mathbb{R}$ is smooth with compact support. Specifically,
\begin{equation}
\pi(u) = \langle u,e_1\rangle, \ldots,\langle u, e_N \rangle.
\end{equation}

The Frechet derivative in a direction $h \in H$ for the {\em cylindrical} test function is:
\begin{equation}
\langle D\varphi(u), h \rangle = \frac{d}{ds}\bigg|_{s=0} \varphi(u + sh) = \frac{d}{ds}\bigg|_{s=0} f(\pi(u + sh)) 
\end{equation}
Since $\pi$ is a linear projection operator:
\begin{equation}
\pi(u + sh) = \pi(u) + s\pi(h),
\end{equation}
we find:
\begin{equation}
\langle D\varphi(u), h \rangle =\frac{d}{ds}\bigg|_{s=0} f(\pi(u) + s\pi(h)) = \langle D_\pi f(\pi(u)), \pi(h) \rangle
\end{equation}
where $D_\pi f$ denotes the gradient of $f$ with respect to its argument in $\mathbb{R}^N$.

By definition of the adjoint operator, we can write:
\begin{equation}
\langle D_\pi f(\pi(u)), \pi(h) \rangle = \langle \pi^*(D_\pi f(\pi(u))), h \rangle
\end{equation}
Therefore, the Fréchet derivative is:
\begin{equation}
D\varphi(u) = \pi^*(D_\pi f(\pi(u)))
\end{equation}

If we now move back to the Liouville equation for measure $\mu_t$, we have, for a cylindrical test function:
\begin{equation}
\langle D\varphi(u), F(u) \rangle = \langle \pi^*(D_\pi f(\pi(u))), F(u) \rangle
\end{equation}
and, using the definition of the adjoint:
\begin{equation}
\langle \pi^*(D_\pi f(\pi(u))), F(u) \rangle = \langle D_\pi f(\pi(u)), \pi(F(u)) \rangle
\end{equation}

Define the marginal measure $\nu_t = \pi_{\#} \mu_t$ on $\mathbb{R}^N$, in which $\pi_{\#}$ is the map from the measure on the infinite dimensional space, to the pdf in the finite dimensional space, to find (using the disintegration theorem):
\begin{equation}
\int_{\mathbb{R}^N} f(\xi) \, d\nu_t(\xi) = \int_H f(\pi(u)) \, d\mu_t(u)
\end{equation}
Then:
\begin{equation}
\frac{d}{dt} \int_{\mathbb{R}^N} f(\xi) \, d\nu_t(\xi) = \int_H \langle D_\pi f(\pi(u)), \pi( F(u)) \rangle \, d\mu_t(u)
\end{equation}

For this to define a Kolmogorov equation on $\mathbb{R}^N$, we identify the drift term $F_{\nu}$ for the pdf $\nu_t$ via:
\begin{equation}
\int_H \langle D_\pi f(\pi(u)), \pi(F(u)) \rangle \, d\mu_t(u) = \int_{\mathbb{R}^N} \langle D_\pi f(\xi), F_{\nu}(\xi) \rangle \, d\nu_t(\xi)
\end{equation}
which defines $F_{\nu}$ as
\begin{equation}
F_{\nu}(\xi) = \mathbb{E}[\pi(F(u)) \mid \pi(u) = \xi]
\end{equation}
The Liouville equation for the pdf now becomes:
\begin{equation}
\frac{\partial \nu_t}{\partial \xi} + \nabla_\xi \cdot (\nu_t F_{\nu}(\xi)) = 0
\end{equation}

\subsection{The Closure Problem and its solution}
The effective drift coefficient is, as derived above,:
\begin{equation}
F_{\nu}(\xi) = \mathbb{E}[\pi(F(u)) \mid \pi(u) = \xi]
\end{equation}
To compute this conditional expectation, we need to know how $F(u)$ depends on the full infinite-dimensional solution $u$, not just on the finite-dimensional projection $\xi = \pi(u)$. To see this more clearly, we write  $u = u_\parallel + u_\perp$ where $u_\parallel = \pi(u)$ and $u_\perp \in H_\perp$:
\begin{equation}
F_{\nu}(\xi) = \mathbb{E}[\pi(F(u_\parallel + u_\perp)) \mid u_\parallel = \xi]
\end{equation}
This requires the conditional distribution $\mathbb{P}(u_\perp \mid u_\parallel = \xi)$, which encodes the full infinite-dimensional dynamics. This is the closure problem: to solve the evolution equation for the pdf on the finite-dimensional space, we need the measure over the infinite-dimensional space.

However, physical systems are causal, meaning that the solution at a point $(x,t+\Delta t)$ depends only on a finite dimensional domain $S_x$ at time $t$. This is related to the finite maximum propagation speed of information in any physical system. For instance, in the shallow-water system described in the main text, the maximum propagation speed is $\sqrt{gH}$, and the domain of dependence will have radius $\Delta t \sqrt{gH}$.

This means that the conditional expectation 
\begin{equation}
F_{\nu}(\xi) = \mathbb{E}[\pi(F(u_\parallel + u_\perp)) \mid u_\parallel = \xi]
\end{equation}
depends only on the conditional distribution of $u_\perp$ within the domain of dependence, i.e., in regions causally connected to the resolved modes $\xi$ over the time step $\Delta t$.

If we now think about an explicit finite difference scheme, the time step is chosen such that the domain of dependence is smaller than the distance between grid points. This is needed to avoid the numerical scheme  becoming unstable. Hence, over one model time step we only need the expectation over that part of $u_\perp$ that is smaller than the grid distance. Of course, one can use an implicit scheme in the model, but the physical dependence on the maximal propagation speed remains, ensuring that the domain of dependence will always be of the order of the grid spacing. 

This argument shows that to solve the evolution of the pdf, we only need that part of the infinite-dimensional space that is in the domain of dependence of each grid point in the finite-dimensional space, in other words, the closure is only local. Furthermore, we know from extensive experience solving PDEs numerically, that the influence of the subgrid-scale physics on the resolved physics can be controlled with advanced numerical schemes and local parameterizations. This suggests that also the influence of the subgrid scales on the evolution of the pdf can be controlled in a similar manner, especially when we consider that we can approximate that evolution via ensemble integrations over the finite-dimensional space. This, then, is a strong argument for why information flow calculations based on pdfs informs us about information flow in infinite dimensional spaces. Specifically, we can approximate, for a certain grid point 
\begin{equation}
F_{\nu}(\psi) \approx F(\psi,\bmpsi_e,t)
\end{equation}
where the latter is obtained from a direct discretization of the PDE, as in the main text.
Most importantly, this causal argument is crucial in understanding why the calculations needed for the different contributions to information flow are local and hence low-dimensional. 

As an aside, parabolic PDEs have infinite propagation speed. A disturbance at any point spreads instantaneously to all other points, resulting in a domain of dependence that contains the whole space. However, it should be remembered that parabolic PDEs are \textit{effective theories} valid at certain scales, and they emerge from averaging or coarse-graining over smaller scales. This is also reflected in the fact that explicit finite difference schemes, with their local domain of dependence, can be highly accurate (but perhaps not most efficient) for parabolic equations. 

Finally, it should be mentioned that if the PDE has a specific linearity or has a point-wise nonlinearity, e.g. a nonlinearity that does not contain derivatives, such as $u^3$, the conditional expectation can be calculated exactly. Assume the drift splits into a differential part and a local nonlinearity,
\begin{equation}
F(u) = A u + g(u),
\end{equation}
where $A$ is a linear operator and $g(u)(y)=g(u(y))$ is pointwise. Then
\begin{equation}
\pi(F(u)) = \pi(A u) + \pi(g(u)),
\end{equation}
Now use for the linear term:
\begin{equation}
\pi(Au)_i = \langle e_i,Au \rangle = \langle A^* e_i,u \rangle
\end{equation}
If $A^* e_i \in \text{span}\{e_1, \ldots, e_N\}$ for every $i$, then $\pi(A u)$ remains a projection to the finite-dimensional space, and there is no closure problem for this term. An example would be if $A$ is a constant matrix. For the pointwise-nonlinear term we find that $g(u)$ is directly projected on the finite-dimensional space.

\section{: Practical calculation of integrals}
To calculate the relative entropy we use \citeA{Ebrahimi1994}. We found that this estimator is more accurate for our purpose than nearest-neighbor estimators.
Assume we have $N$ samples $\psi_i$ sorted from low to high, the differential entropy estimator is, following \citeA{Ebrahimi1994}:
\begin{equation}
H = \frac{1}{N} \sum_{i=1}^N \frac{N}{c_i m} (\psi_{i+m}-\psi_{i-m}),
\end{equation}
in which
\begin{equation}
c_i = \left\{
\begin{matrix}
1+i/m &\;\;\;{\rm for} \;\;\; 1 \leq i\leq m\\
2 &\;\;\;{\rm for} \;\;\; m+1 \leq i\leq N-m\\
 1+(m-i-1)/m &\;\;\;{\rm for} \;\;\; N-m+1 \leq i\leq N\\
\end{matrix}
\right. ,
\end{equation}
and $m=\sqrt{N}+0.5$. Furthermore, $\psi_{i+m}-\psi_{i-m}=\psi_{i+m}-\psi_1$ for $i \leq m$, and $\psi_{i+m}-\psi_{i-m}=\psi_{N}-\psi_{i-m}$ for $i \geq N-m+1$.

The information estimator based on the relative entropy is obtained from this, assuming a reference density $q(\psi) = N(m_{ref},v_{ref})$, as:
\begin{equation}
I_{\psi} = - H + \log\bigl(\sqrt{2\pi v_{ref}} \bigr) -\frac{1}{N} \sum_{i=1}^N \frac{(\psi_i-m_{ref})^2}{2 v_{ref}}.
\end{equation}
It was found that using the analytical expression for $q$ in the information estimate was more accurate than an ensemble estimate.
We can use this algorithm also for expressions 
\begin{equation}
G_{\psi} = \int g(\psi) p(\psi)\log \frac{p(\psi)}{q(\psi)} \;d\psi ,
\end{equation}
by simply adding the factor $g(\psi_i)$ under the summations.

We also need to estimate integrals of the form 
\begin{equation}
\int p(\psi,\bmpsi_e) g(\psi) \Dp{}{\psi}\Biggl(\log \frac{p(\psi)}{q(\psi)}\Biggr)\; d\psi d \bmpsi_e .
\end{equation}
For this we use an ensemble estimate of $d/d\psi \log p(\psi)$, sometimes (confusingly) called the score. The methodology is largely follows Pinkse and Schurter (2020) and is kernel based. To find the kernel we can use that, for a Gaussian, $d/d\psi \log p(\psi)$ is a straight line, resulting in a quadratic kernel as explained below.

Assume the samples are drawn from an unknown pdf $p(z)$. We will use the identity 
\begin{equation}
\frac{d}{dz}p(z) = p(z) \frac{d}{dz} \log p(z) ,
 \end{equation}
in the following.
We will find a local estimate of the gradient of the log via a partial integration involving a kernel $g(t)$. We first define the variable
\begin{equation}
t = \frac{z-x}{h},
 \end{equation}
in which $z$ and $x$ are scalar samples and $h$ is a bandwidth. The kernel function is forced to satisfy $g(-1)=g(1) = 0$. 
We then note that
\begin{eqnarray}
\int_{x-h}^{x+h} \frac{d}{dt} \left(g(t) \right) p(z) \; dz & = & 
h\int_{x-h}^{x+h} \frac{d}{dz} \left(g\left(\ \frac{z-x}{h}\right) \right)p(z) \; dz \nonumber \\
& = &  \left. h g\left(\ \frac{z-x}{h}\right) p(z) \right|_{x-h}^{x+h} -  
h\int_{x-h}^{x+h} g\left(\ \frac{z-x}{h}\right)  \frac{d}{dz}p(z) \; dz \nonumber \\
& = &  -h\int_{x-h}^{x+h} g\left(\ \frac{z-x}{h}\right)  p(z) \frac{d}{dz}\log p(z) \; dz,
 \end{eqnarray}
where we used the conditions on the kernel to eliminate the constant term.  We now define the average gradient of the log prior over the interval as 
\begin{equation}
\int^{x+h}_{x-h}  \frac{d}{dt} \left(g(t) \right) p(z) \; dz = - \frac{d}{dx} \log p(x) \int^{x+h}_{x-h}  g(t)  p(z)h  \; dz.
 \end{equation} 

We now use a sample estimate of the pdf, so we write
\begin{equation}
p(z) = \frac{1}{N} \sum_{i=1}^N \delta(z - z_i),
 \end{equation}
and define $i \in a_x$ as those samples that are within the interval $[x-h,x+h]$. We can now define the average gradient of the log over the interval $[x-h,x+h]$ via:
\begin{equation}
\frac{d}{dx} \log p(x) = -\frac{1}{h}\frac{ \sum_{i \in a}\frac{d}{dt} g(t_i)} {\sum_{i \in a} g(t_i)}.
 \end{equation}

We choose the kernel function
\begin{equation}
g(t)= (t+1)(1-t) = 1-t^2 \;\;\;\;\; {\rm on} -1<t<1,
 \end{equation}
and zero outsize this interval. The reason for this choice is that it fulfills $g(1)=g(-1)=0$ and its derivative is 
\begin{equation}
\frac{d}{dt}g(t)=  - 2t \;\;\;\;\; on -1<t<1,
 \end{equation}
which is linear in $t$. This facilitates that for Gaussian- or exponential-tailed pdfs the gradient is indeed linear. With this kernel we find:
\begin{equation}
\frac{d}{dy} \log p(x) = \frac{1}{h}\frac{ \sum_{i \in a} 2t_i } {\sum_{i \in a} (1-t_i^2) },
 \end{equation}
in which
\begin{equation}
t _i= \frac{z_i-x}{h}.
 \end{equation}

After trial and error, the bandwidth is chosen as
\begin{equation}
h =2 \sigma_z ,
 \end{equation}
in which $\sigma_z$ is the standard deviation of the base points.
Then it is checked if there are at least 10 base points within a distance $h$ from the test point we are interested in. If that is not the case we increase the bandwidth by a factor 2 and test again, until the condition is fulfilled.

%
%

\section*{Open Research Section}
The data can easily be reproduced using the shallow-water model originally developed by Robin Hogan, available at \begin{verbatim} https://www.met.reading.ac.uk/~swrhgnrj/shallow_water_model/ \end{verbatim}

\acknowledgments
This work was initiated via funding through the European Research Council project CUNDA (No. 694509). The author thanks an anonymous reviewer and Dr. S. Liang for the deep comments that led to a substantially improved paper.

\noindent The author declares no conflict of interest.
%
%

\bibliography{references}

%
%
%
%
%

\end{document}